# Battery Lifetime Prediction using Data-driven Modeling Approaches

Vikram C Patil, *North Carolina State University, Raleigh, USA*.

**Abstract**

Batteries are ubiquitous today, with applications ranging from smartphones, watches, and laptops to electric cars, drones, and electric aircraft. Lithium-ion batteries are widely used in these applications due to their high energy density, rechargeability, and low lifecycle cost. Understanding the lifetime of lithium-ion batteries is essential for their effective utilization across many domains. In this study, data-driven modeling approaches are explored to predict the lifetime of lithium-ion batteries using various measurable battery parameters. A battery dataset from NASA's electric aircraft experiments was used, which included 17 predictor variables and remaining flight time as the response variable representing battery lifetime. The dataset contained more than 4,000,000 rows. However, the original dataset provided limited directly useful information about battery utilization over time; therefore, feature engineering was performed to generate more informative variables. Additionally, dimensionality reduction using principal component analysis (PCA) was applied to reduce computational cost and model complexity by selecting a smaller number of principal components as predictors for model development. Random forest and neural network models were explored for battery lifetime prediction using the engineered features. Multiple neural network configurations were evaluated, including single- and double-hidden-layer architectures with varying numbers of nodes. Mean squared error (MSE) on the test dataset was used as the performance metric for model comparison. The results indicate that data-driven modeling approaches are effective for battery lifetime prediction, with neural network models outperforming other models based on the MSE metric. Furthermore, neural networks demonstrate robustness in handling high-dimensional battery data. Overall, the complex behavior of batteries can be effectively modeled using data-driven machine learning techniques to predict battery lifetime with reasonable accuracy. Future research in this area could further enhance prediction accuracy, computational efficiency, and model reliability.

# 1. Introduction

Batteries especially rechargeable Li-ion batteries are ubiquitously used for a range of applications like smartphones, laptops, drones, e-scooters, electric cars, buses, aircrafts, etc. (Pistoia, G. 2013). They also play an important role in curbing global warming and climate change by reducing the use of fossil fuels and by increasing the use of more renewable energy resources (Climate Central, 2019). The characteristics like high energy density and long lifetime of Li-ion batteries compared to other battery types make them preferable for many applications (Yoshio, M., Brodd, R.J. and Kozawa, A. 2009). For many applications of Li-ion batteries, lifetime prediction of the battery is crucial for reliable use and evaluating financial viability. For example, car manufacturers or owners of electric cars would like to know the longevity of the battery used in the car. If battery life is significantly lower than the car's life, then battery replacement needs to be planned. Conversely, a battery lasting longer than a smartphone may be unnecessary and lead to the overburden cost to the owner. Therefore, lifetime prediction Li-ion batteries for a variety of battery applications could be highly valuable for their appropriate use in a variety of applications. Moreover, understanding uncertainty in the lifetime prediction could be highly useful for determining the warranty time for these batteries. With quantified uncertainties or probabilities in the lifetime prediction, manufacturers of Li-ion batteries can mitigate the financial risks of warranty claims due to battery failures by providing a warranty with less probability of failure.

There are various approaches that can be used for modelling the lifetime of batteries in general. Figure 1. (Hosen et al. 2021, pp. 102060) illustrates three major types of approaches that can be used with their features. The first one is the electrochemical model also called physics-based modeling because it is based on physical laws governing various mechanisms of battery aging over its use. This physics-based approach can result in a highly accurate model with the use of limited experimental data; however, it is complex, computationally expensive, and less applicable for real-world conditions. The second approach is a semi-empirical modelling approach that uses less complex mathematical equations in simplified forms governing equations describing the physics in the battery. The semi-empirical models generally need good quality and a reasonable quantity of experimental data with diverse application conditions. The data is used to tune the parameters in the model. Semi-empirical models are computationally easy compared to electrochemical models and they can show significantly lower model accuracy compared to electrochemical models. The third major modelling approach that has actively been explored with the progress of the data science field is data-driven modelling. In the data-driven modelling approach, advanced data science techniques are used to model battery aging and lifetime. Data-driven modelling requires limited knowledge of physical mechanisms in batteries but can build highly complex models that can be used for a range of applications with high accuracy. Data-driven models are proven to be easy to develop but generally require a large amount of data covering a range of operating conditions.

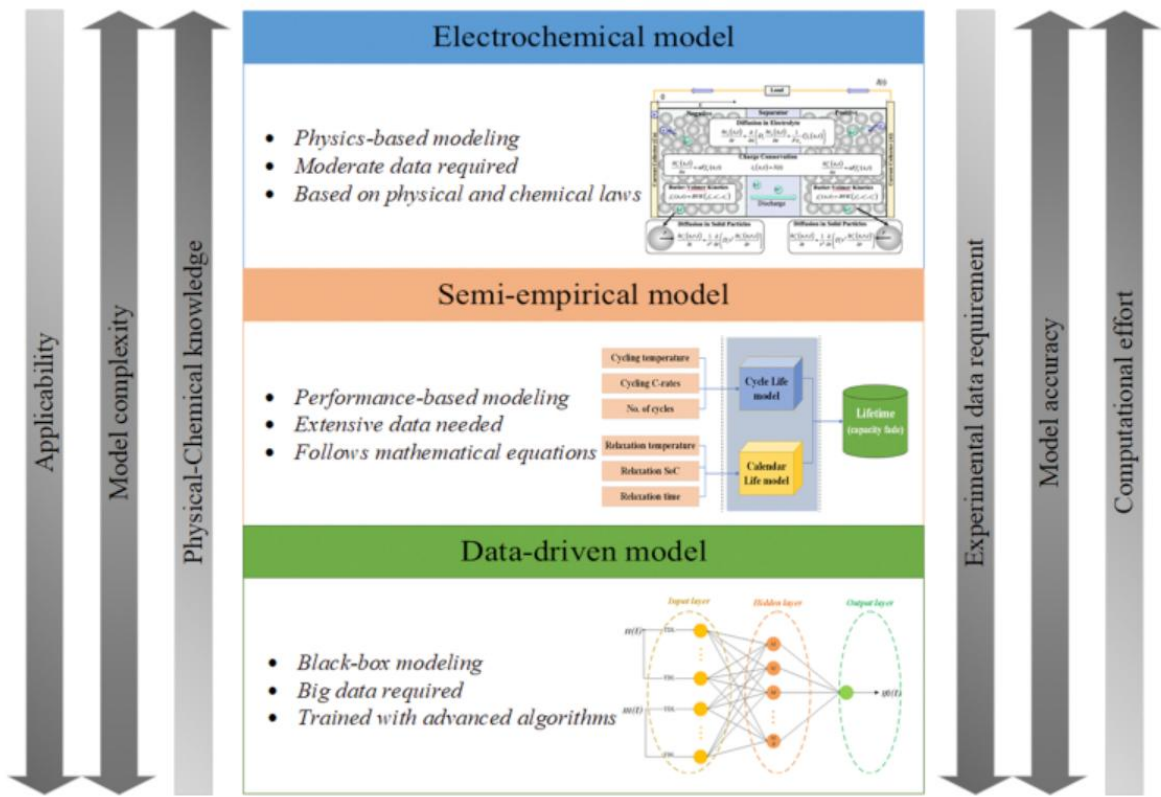

**Figure 1.** Battery lifetime modelling approaches at glance. (Source: Hosen et al. 2021, pp. 102060)

With technological advancements and more awareness about renewable energy, electric vehicles are becoming increasingly common. However, the dream of an electric commercial aircraft is still far from reality. There are many challenges that are presented when a commercial aircraft is powered using an electric propulsion system, for example, the absence of a practical cooling system, the weight of the engines, etc. Yet the major limiting factor is the battery, especially estimating the remaining battery operating time and subsequently the remaining flying time of the aircraft. It is imperative to predict the remaining flying time to alert the flyer to initiate the landing procedure before the battery's State of Charge (SOC) is two minutes away from falling below the safe 30% threshold. There is ongoing research at the National Aeronautics and Space Administration (NASA) to address this issue. In this project, battery data on electric aircraft is considered to predict the lifetime of the battery using data-driven machine learning models.

## 1.1. Aim

Develop and identify suitable data-driven machine learning models for predicting the lifetime of Lithium-ion batteries using on-field operation data on batteries.

## 1.2. Objectives

i) Analyze and identify suitable features from the battery lifetime data useful for predicting the lifetime with reasonable accuracy.

ii) Develop and test various machine learning models to predict battery lifetime.

iii) Compare and evaluate the effectiveness of various models on different model assessment metrics.

iv) Assess the effectiveness and limitations of various machine learning models in predicting the lifetime of the battery.

v) Identify and evaluate possible uncertainty estimation techniques suitable for the data-driven battery lifetime modelling approach.

## 1.3. Research Questions

i) Is data-driven modelling an effective approach for predicting the lifetime of a battery with reasonable accuracy?

ii) Are there any machine learning models effective in predicting the lifetime of a battery outside the conditions covered by the training dataset?

iii) Can data-driven models provide a measure of uncertainty in battery lifetime prediction?

## 2. Literature Review

The data-driven modelling approach was seldom used for battery modelling in past due to the limited availability of data, the higher cost of generating data, and the difficulty in interpreting the data-driven models. Many earlier studies on battery lifetime modelling were based on a semi-empirical modelling approach using experimental data generated on batteries covering different operating conditions. An overview of different approaches to battery lifetime prediction is presented by Zhang, Liang, and Zhang (2017, p. 012134).

Machine learning techniques have been explored recently for the application of lifetime prediction of Li-ion batteries. Severson et al (22019, pp. 383-391) used a data-driven modelling approach on data before significant capacity degradation to predict battery cycle life. In Severson et al. work, discharge voltage curves from early cycles were used to predict and classify battery cells by cycle life. Features created by linear and non-linear transformations of raw data were used to train Elastic net. The model is shown to be effective in predicting the cycle life of the battery with a test error of 9.1% on 124 commercial lithium-ion battery data.

Another study by Zhu, Zhao and Sha (2019, p.e98) used data recorded on 138 LFP/graphite battery cells. They constructed a classification model to predict battery cells into two distinct classes 'low lifetime' and 'high lifetime'. Various machine learning methods like decision trees, K-nearest neighbors, Neural Net, AdaBoost, GPR, Random Forest, Support Vector Machines, and Naïve Bayes were fitted and compared for better classification accuracy. Figure 2 compares the prediction accuracy of various machine learning methods on the same dataset observed in their study. Many techniques have shown a good accuracy and the decision tree model showed the best test accuracy metric. Although the method has shown effectiveness in classifying the battery cells into low or high-life classes with reasonable accuracy, the method lacks in providing prediction of battery life in years.

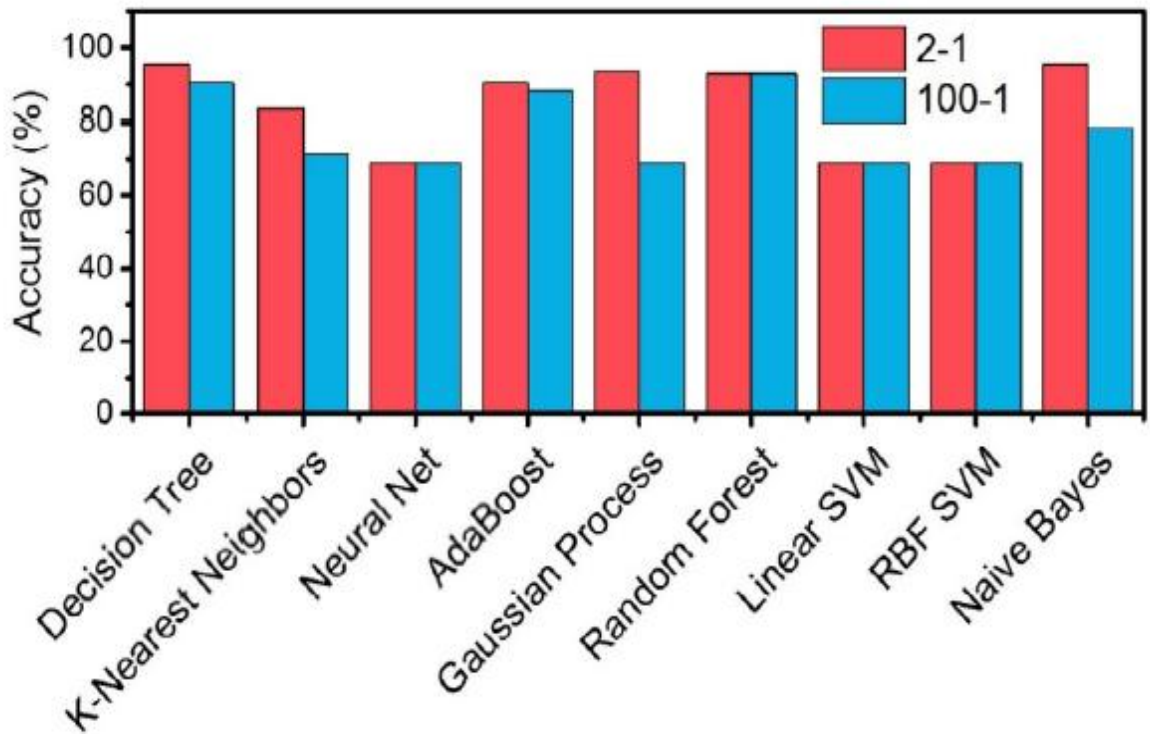

**Figure 2.** Prediction accuracy comparison of various machine learning algorithms on the same test dataset. The legend "2-1" and "100-1" indicate two sets of feature combinations. (Source: Zhu, Zhao and Sha 2019, p.e98).

In a recent study, Hosen et al. (2021, pp. 102060) compared various battery lifetime modelling approaches. In their study, two machine learning-based models based on Gaussian Process Regression (GPR) and Artificial Neural Networks (ANN) were compared with a semi-empirical model. GPR model with exponential kernel was created whereas a nonlinear autoregressive dynamics network with external inputs (NARX) ANN model was used. The model assessment indicated that the NARX ANN model was the best model for Root mean squared error (RMSE), Mean absolute error (MAE), and computational time metrics. Importantly, both the data-driven machine learning modelling approaches had shown higher accuracy than semi-empirical.

In another recent study, a comprehensive Machine Learning (ML) based framework consisting of feature extraction, feature selection, and ML-based prediction of battery life from early cycles data was proposed by Fie et al (2021, p. 120205). Six ML-based models including elastic net, Gaussian Process Regression (GPR), Support Vector Machine (SVM), Gradient Boosting (GB) regression tree, random forest, and neural network were considered on a set of 42 features extracted from only the first 100 cycles. It was observed that elastic net, GPR, and SVM models showed better test results than the GB-regression tree, random forest, and neural network. Moreover, wrapper-based feature selection has shown effectiveness in improving the prediction performance for all the models. Although many previous studies have showcased the effectiveness of data-driven models for battery lifetime prediction, many of these methods have not proven their credibility for predicting the operating conditions outside the data used for model development. Lifetime prediction from limited data outside operating conditions of battery training data is needed for the fast-paced battery industry.

Moreover, the topic of data-driven lifetime modelling of Li-ion batteries is of interest to many researchers working on the advancement of battery technology. National Renewable Energy Laboratory of the USA is working on a physics-based machine learning approach for improving lifetime prediction accuracy (U.S. Department of Energy, 2022). Additionally, Argonne national laboratory (ANL) of the USA is developing a comprehensive AI-driven approach to predict battery life and degradation trends with limited usage of data (Argonne National Laboratory 2022). For the growing battery usage with many applications demanding a longer battery lifetime, prediction of battery lifetime accurately is of significant interest to the battery industry and academics.

## 3. Dataset

There are limited battery datasets available publicly. An article (BatteryBits, 2020) reviews and compares fourteen publicly available battery datasets from various sources. The cycle life prediction dataset from Stanford university is a good fit for testing data-driven approaches. Figure 3 (BatteryBits, 2020) shows capacity aging data of 135 Li-ion battery cells tested under different operating conditions from this dataset. The lifetime of a Li-ion battery is represented either in time or the number of cycles to reach the capacity of the battery to a certain percentage (generally 70% or 80%) of the initial capacity. The capacity of the battery reduces as it is used or rested. The black dotted line in the plot is the lifetime condition to reach 80% of initial capacity and solid circles denote the end-of-life time for the individual plot. It can be observed that in this dataset lifetime ranges from a few hundred cycles to a few thousand cycles. These differences can be due to the different operating conditions the battery experiences when in use. This dataset only provides cycle life information for the highest number of cells (135 total) of any publicly available dataset with rich information about features. It is important to get on-field data from the battery in the application to get a rich dataset with real-world data with rich information.

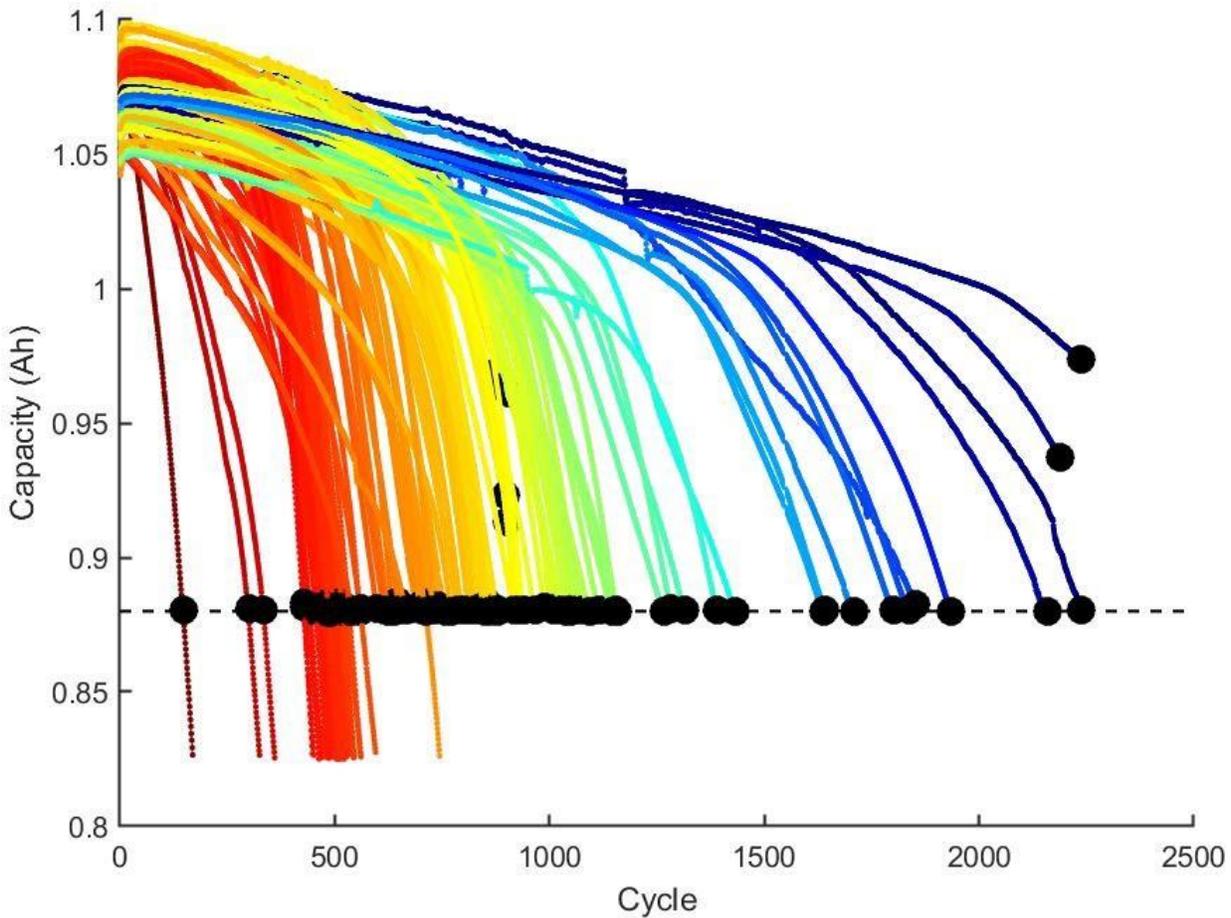

**Figure 3.** Capacity aging of 135 Li-ion battery cells tested under different operating conditions. (Source: BatteryBits, 2020)

The data from various sources was attempted to access to get it in a useful format. However, many of battery data is available in encoded formats which makes it difficult to use. Moreover, only a few sources shared a detailed description of data variables. It was very troublesome to get data with details including battery lifetime. One unique dataset that was available easily but required a lot of data processing to get in the right format was from NASA's prognostic center of excellence laboratory.

Electric aircraft generally record a significant amount of detailed battery data. Battery data recorded from electric aircraft not only have rich information but also highly real-world data. The application demands higher safety standards and therefore the data quality is generally of a high standard. Therefore, data from electric aircraft was considered for this project. Specifically, the data from ground test experiments conducted by NASA on Electric Edge 540 aircraft (NASA Prognostics Data Repository). These are aerobatic aerial vehicles, and their sub-scale version was used for these experiments. There is a large amount of data available for these experiments. For our analysis, we have used data from 9 experiments, which is approximately 4,000,000 rows.

## 3.1. Variables

The dataset consists of 18 variables describing three major types of measurements included:

i) Remaining flying time estimate for the aircraft – This variable is battery lifetime and therefore it is the response variable for our analysis. It is measured in seconds.
ii) Aircraft states – Three variables namely 'Revolutions per minute (RPM)', 'Forward Motor Controlled Sensor' and 'After Motor Controlled Sensor measure' the aircraft state at a given point of time.
iii) Battery SOC estimates – There are 14 variables describing the state of charge (SOC) of different batteries used in the aircraft. Battery's state of charge has been described by measures of its voltage, current, and temperature at a given point in time.

These variables are generally measured and stored as a part of Battery Management systems recordings. These variables include both directly measured and indirectly calculated variables. The response variable 'Remaining flying time' and aircraft states are directly measured. Battery SOC estimates are calculated variables from direct measurements of voltage, current and temperature of the battery. The direct variables generally do not have significant measurement errors. The indirect variables many times have both measurement errors due to direct measurement variables used to calculate the indirect variable and errors due to estimation.

The list of names of all the 18 variables is as below.

**Response Variable:**

- Remaining flying time estimate i.e. battery lifetime (in seconds)

**Variables describing Aircraft state:**

- Revolutions per Minute (RPM)

- Forward Motor Controller Sensor (FMC)

- AMC – After Motor Controller Sensor

**Variables describing Battery SOC:**

- LLF20V Lower Left Front – Battery Voltage

- ULA20V Upper Left After – Battery Voltage

- LRF40V – Lower Right Front – Series Combined Voltage

- URA40V – Upper Right After – Series Combined Voltage

- LRF20V – Lower Right Front – Battery Voltage

- URA20V – Upper Right After – Battery Voltage

- LLF20C – Lower Left Forward – Battery Current

- ULA20C – Upper Left After – Battery Current

- LRF40C – Lower Right After – Series Combined Current

- URA40C – Upper Right After – Series Combined Current

- LLF20T – Lower Left Front – Temperature

- ULA20T – Upper Left After – Temperature

- LRF40T – Lower Right Front – Series Combined Temperature

- URA40T – Upper Right After – Series Combined Temperature

# 4. Methodology

The available dataset was split into training, testing, and validation datasets to train, test, and validate models. Various prediction machine learning models were trained on the training dataset to predict the cycle life of the battery. First simplistic models like regression, decision tree, and single layer small Neural Network, are tested to understand the effectiveness and accuracy of the simplistic model in predicting battery lifetime. It was observed that simplistic models are not effective, and the prediction was not better than a random prediction. Furthermore, complex models including Gaussian Process Regression, Random Forest, Deep learning, etc. are tried to check the possibility of improving model accuracy with complex machine learning models. However, it was observed initially that complex machine learning models were also not that effective in predicting battery lifetime.

After many unsuccessful models and prediction attempts, it was come to notice that feature selection needs to be performed. Correlation analysis among the various input variables was explored. It was observed that many of the variables were redundant and had high correlations. Moreover, limited features were present which can provide useful information needed for better prediction. Therefore, feature selection was explored extensively to identify the influencing variables. Few of the aircraft state variables needed to be preprocessed to include an overall history of battery usage than just instantaneous measurements. Therefore, features having cumulative state variables are derived to provide an assessment of the history of battery usage. Furthermore, the possibility of reduction of non-influencing features was explored using principal component analysis (PCA). The dataset was huge with more than 4,000,000 rows and 18 variables. Feature selection was observed to highly important process in data-driven machine learning modeling on battery datasets. Additionally, the possibility of generating and using a synthetic dataset and its effectiveness was tested for battery lifetime prediction. However, it was observed that a synthetic dataset was not necessary as the original dataset was large enough to model the battery lifetime reasonably well.

After feature selection, random forest and deep learning models started showing a prediction with better accuracy. These models were further tuned to get optimized accuracy from each of these models. Finally, prediction errors from both models are compared on a new testing dataset based on performance comparison metrics like MSE, absolute error, and computational time to identify the best modelling framework suitable for battery lifetime prediction.

The following is the list of anticipated outcomes using this methodology from this project.

i) Important features which affect the lifetime of the battery. This information provides insights into identifying useful datasets and reducing computational time by reducing unnecessary information. Moreover, this would guide generate new data by designing experiments including important input features to build better models in the future.

ii) Various data-driven models for predicting lifetime and their comparison on a common metric on unseen data.

iii) The best-performing model would be used for predicting the lifetime of the battery. For a new input feature specifying battery conditions, the machine learning model would provide an estimate of a lifetime.

### 4.1. Ethical Considerations

The dataset used is openly available on NASA's website under the NASA prognostics data repository. The source of the dataset is cited in the reference. Moreover, the dataset contains no information which could violate ethical norms. Any bias in the dataset was avoided by randomly selecting battery files among many and checking any inconsistency in the variables. There was no inconsistency or bias observed in the variables of the dataset. Moreover, during the feature selection process, it was observed that the same features were identified on different cells and no bias was observed in the dataset. The feature engineering was performed on state variables also performed to make sure the resultant variable were without any additional bias. During the project execution, ethical considerations were considered and all the information available is presented appropriately. Additionally, the open assessment computational libraries were used for project implementation. The personal computing platform was used for computing purposes without harming any entity or person. The results of the projects are communicated as the observed and detailed analysis was performed based on the available information from this project data.

## 4.2. Challenges

The lifetime of batteries is influenced by a variety of factors in a non-linear complex form which makes it challenging to model. There are very limited datasets available publicly for battery lifetime because testing batteries is a costly and time-consuming process. Moreover, due to competitive, copyright, and license restrictions, many battery manufacturers do not make available or publish lifetime data publicly. Therefore, there is no diverse and large data available publicly. Moreover, available data also needs to have many important features. The battery is governed by highly complex mechanisms and to capture most mechanisms in the data-driven model, a rich dataset including high feature details may be required. Additionally, there are a few erroneous and missing values in the dataset. Handling these is challenging as removing these data could further reduce the amount of data. Furthermore, data with replicates is valuable for quantizing uncertainty in perdition including sample-to-sample variation. However, replicates are not available therefore uncertainty evaluation could not be performed.

The dataset used is a real-world battery dataset with many raw measurements which may not have a direct influence on battery lifetime. This was observed when model fitting on raw data was not showing good fit and prediction without any feature engineering. The challenging part was to perform feature engineering to identify features in a form that provides information suitable for model fitting to predict battery lifetime. After an extensive understanding of input and response variables properties, an appropriate feature engineering method was identified to create new derived input variables which can account for the history of operation of the battery. This was crucial as battery lifetime is not just dependent on instantaneous variables but also depend on history of various variables.

# 5. Results and Discussion

## 5.1. Data Preprocessing

The dataset available from the source was only available as a separate file for each battery test. Additionally, variables were not stored in an organized table or dataframe format. When data was loaded, it was observed that data was loading in dictionary format. This was because the data in the raw format was available in the mat file format. Moreover, the variables associated with battery state of charge estimates were in a separate structure. Therefore, data preprocessing was necessary to convert this unstructured data into a structured table or python dataframe format. In data preprocessing, battery state of charge variables were appended to the other variables to have all the data in the table format. This was performed for each data file available for each battery experiment. The data in table format was then converted into a CSV file using a Matlab script. CSV files for each experiment were used to access data and do further analysis in a python environment. As the data file for each experiment was loaded and combined resulting in a significantly larger size with more than 4,000,000 rows and 18 columns.

## 5.2. Feature Engineering

When data was used in their original form, poor model fitting was observed. Therefore, it was noticed that the dependent variables in their original form do not provide valuable information for a good model building. Feature engineering was needed to convert the data variables into a form that include valuable information about the state of battery lifetime. The original data format had instantaneous information about battery usage, but battery lifetime is dependent on the history of usage. Therefore, it was necessary to convert instantaneous information about battery usage into variables that can have information about the history of battery usage.

Since the response variable in this dataset is dependent on functions of the predictor parameters, functional data analysis (FDA) seems like a natural option. FDA is a collection of statistical techniques specifically developed to analyze curve data (Frøslie, 2013). However, Yogev (2014) mentions that it is a common practice to use simple summary measures, such as the area under the curve (AUC) to obtain information from the functions. Therefore, for this analysis estimated AUC for independent variables was used to predict the remaining battery life.

The notion behind using AUC to determine remaining battery life is that all the predictors must be a function of power and responsible for battery consumption over time. Since the area under the power-time graph provides an estimate of energy consumption, the area under the predictor-time curve will yield the proportion of lithium ions consumed over time in the battery. Applying this methodology to all the predictors provides information about the number of Lithium ions consumed over time from the battery.

Thus, feature engineering was performed on the parameters by taking their AUC i.e., cumulative sum, and used them as subsequent variables for prediction modeling. Figure 4 shows the plot for the original independent variable Revolutions per minute (RPM) over time. It can be observed that RPM values vary significantly ranging from 0 to as high as about 5000. The higher the RPM more the Lithium-ion loss during operation and this instantaneous variable value is not a useful indicator of lifetime when used as it is. This RPM variable when converted with an AUC summary measure to account for the history of the variable, the variable is highly helpful for lifetime prediction as Lithium-ion consumed is represented by an AUC summary measure well. Therefore, the cumulative sum of RPM over time is a newly created variable from this feature engineering. Figure 5 depicts the cumulative sum of RPM over time for the same instantaneous RPM data shown in figure 4. It can be observed that cumulative RPM indicates the area under the curve for instantaneous RPM and therefore provides a summary measure useful for battery lifetime prediction.

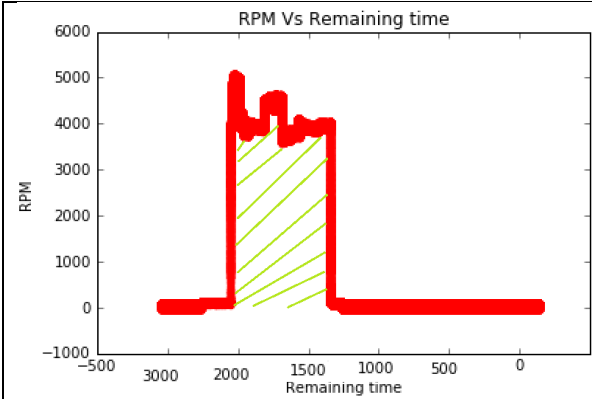
**Figure 4:** RPM response over time

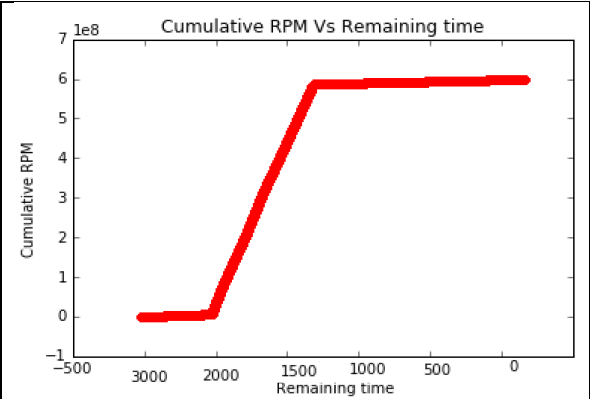
**Figure 5:** Cumulative sum of RPM over time

## 5.3. Correlation Analysis

Since many predictors are functions of voltage, current, and temperature; a high correlation was observed among the parameters. Figure 3 shows the correlation plot among different predictor variables. The dark blue color indicates a high positive correlation whereas the dark red-brown color indicates a high negative correlation. There are 17 predictor variables in the data, and we can observe that many predictors are highly correlated with each other. The correlation among the predictor variables is primarily positive. This is expected as many times an increase of one variable results in the increase of other in battery systems. For example, high RPM is a result of high current withdrawn from the battery. It can also be noted that all the predictors are positively correlated. There is a high correlation between the first seven predictors and also between the remaining ten predictors. Therefore, two major clusters in the predictor variables have been identified which can be used for dimensionality reduction.

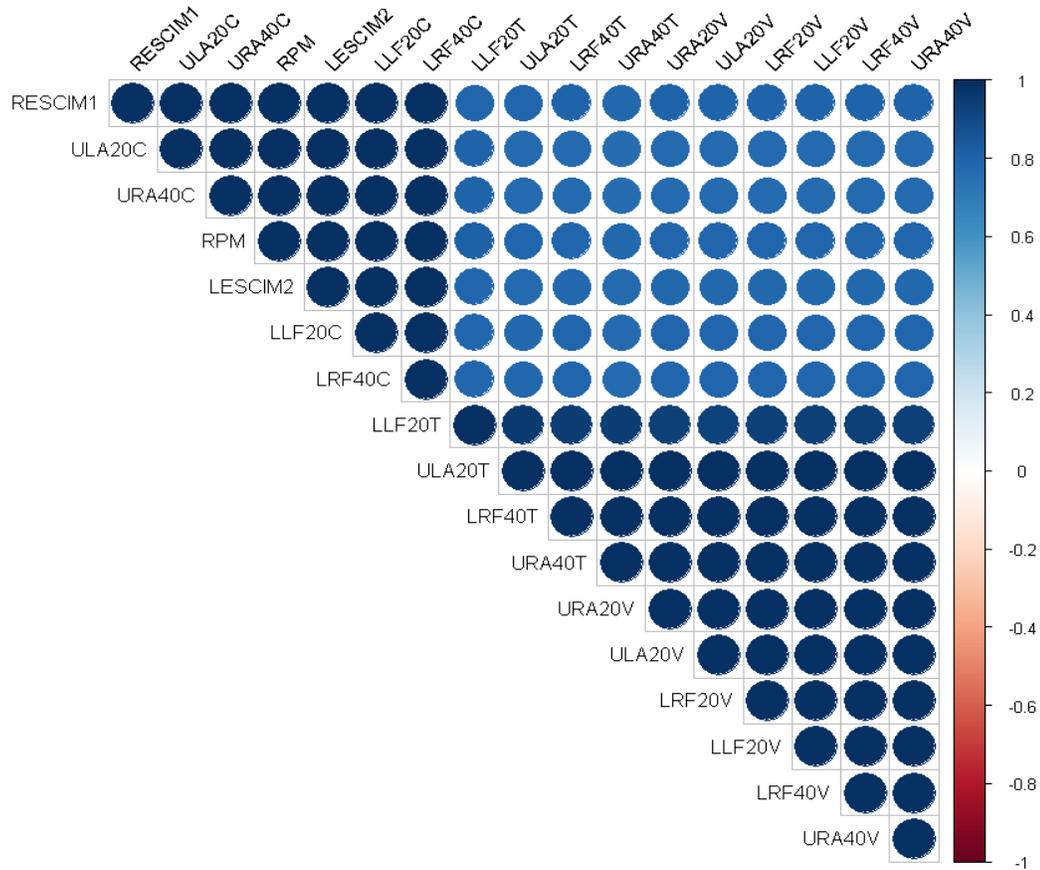

**Figure 6:** Correlation plot between covariates.

## 5.4. Principal Component Analysis (PCA)

Principal component analysis (PCA) is a powerful technique used to reduce dimensionality when predictors are highly correlated. Using PCA, it is possible to keep the same information from the many variables with a significantly lower number of variables. Reduction of dimensionality was necessary to reduce the computational time of machine learning models as more predictor variables demand estimation of more model parameters which is time-consuming during model training. Moreover, reduced dimensionality of predictor variables also makes the model less complex with less number of model parameters.

To reduce the dimensionality of the dataset, PCA was performed to determine the amount of variance that can be captured with a lower number of parameters. Figure 4 shows an elbow plot of the PCA results. We can see that the first two principal components captured 99% of the variance. Any additional PCA variable increases variance explained marginally. The noise in the data also contributes to the variance and therefore it is highly recommended to have significant improvement in variance explained when an additional PCA variable is considered. Clearly, only 2 PCA variables provide 99% of the variation in the dataset. Therefore, the dimensional of predictor variables can be reduced to just 2 transformed variables using PCA without losing much information in the dataset. Since the dataset used has more than 4,000,000 data points, the dimensionality reduction from 17 predictor variables to just 2 variables by PCA means a significant reduction in computational cost.

If 5 principal components are considered, then 99.999% of the variation in the dataset was captured. Therefore, 17 variables can be reduced to just 5 PCA variables without loss of information from the original predictor variables. Additional 3 PCA variables just provide information about 1% of the variation in predictor variables, however, for the predictive models where additional few predictor variables do not contribute significant computational cost, 5 PCA variables can be considered.

PCA is a powerful dimensionality reduction technique that helped reduce computational time significantly for the model training process. However, extra pre-processing was necessary to convert original predictor variables to PCA variables as PCA variables are linear combinations of the original predictor variables. Moreover, the original predictor variables to PCA variable transformation need to be kept the same on the testing dataset for model testing purposes.

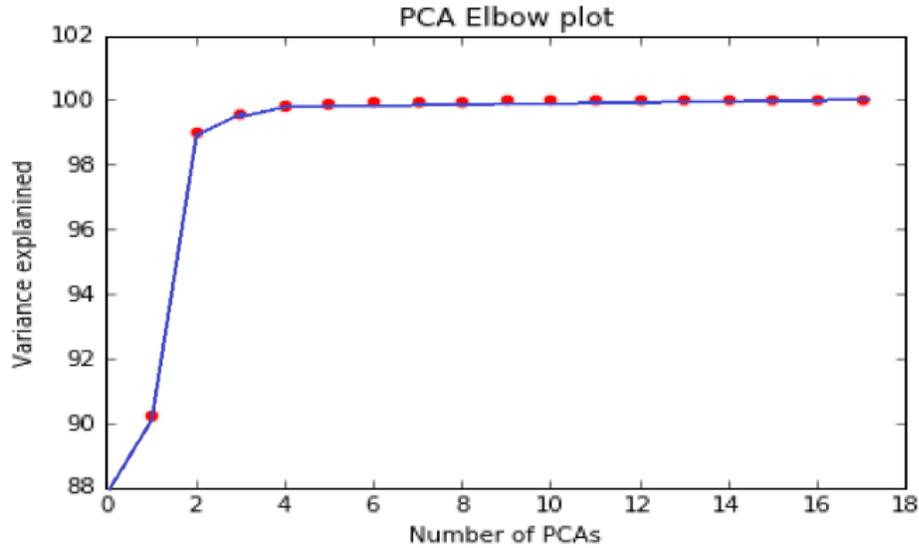

**Figure 7:** Cumulative variance explained by number of principal components.

## 5.5. Predictive modeling

After performing feature engineering and PCA, the dataset was prepared for applying predictive algorithms. The distribution of the response variable is shown in Figure 8. This distribution is unique as it has positive values with uniform distribution for the initial range and exponential distribution at higher values. Clearly, generalized linear models of known distributions may not be suitable for modeling this response. Additionally, from the scientific understanding of battery and flight technologies, it is expected to have a non-linear relationship between covariates including many significant higher-order interactions. Predictive models which can capture non-linear complex relationships with high-order interactions were needed for successful modeling of battery lifetime. Therefore, for data-driven predictive modeling, two prediction models based on Random Forest and Neural Networks were explored. These two machine learning models are the most versatile and robust predictive models. Moreover, both models are known for capturing complex non-linear relationships well. Additionally, real-world battery data might have a lot of noise which can be handled by these models well with the use of supervised learning.

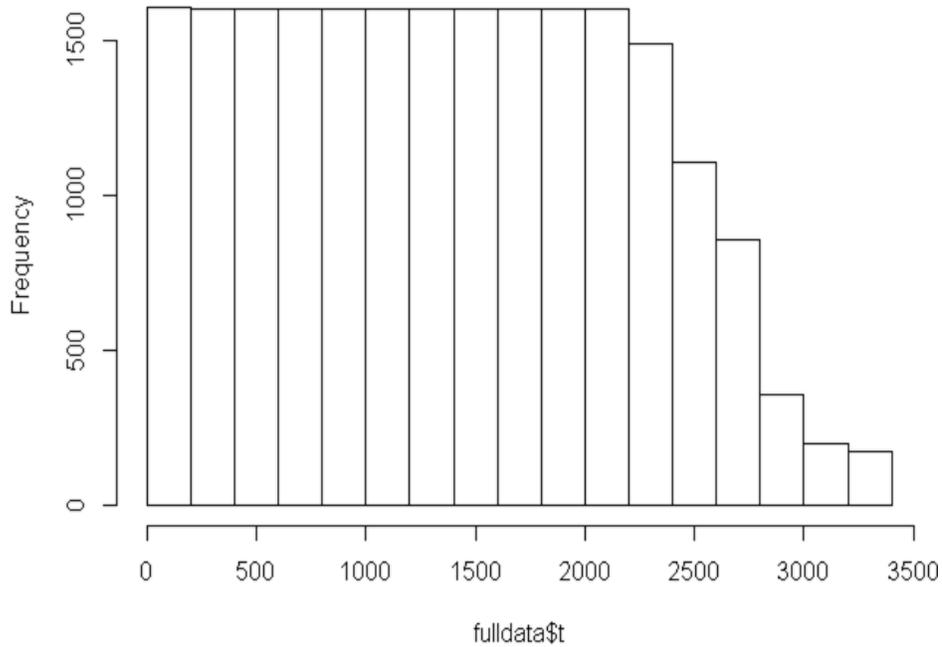

**Figure 8:** Histogram of remaining flight time response.

### 5.5.1. Random Forest

Random forests also referred to as random decision forests are an ensemble supervised learning model which uses multiple decision trees to model fitting. The remaining useful life is a continuous variable therefore random forest regression model was used for fitting model fitting and training on predictor variables. To reduce computational time, only 2 PCA variables are considered as predictor variables with a remaining useful lifetime as a response variable for model tuning. For model training, 50 decision trees were considered on 2 principal components. Additionally, data from 8 experiments was the training dataset, and data from 1 experiment was considered the test dataset. This model performed well on the training set, with an MSE of 27431, however, it showed some deviation from the actual values on the test data. The prediction vs actual response variable fit on test data is shown in Figure 9. Clearly, a significant deviation is observed from the actual response variable in the model prediction values for test data.

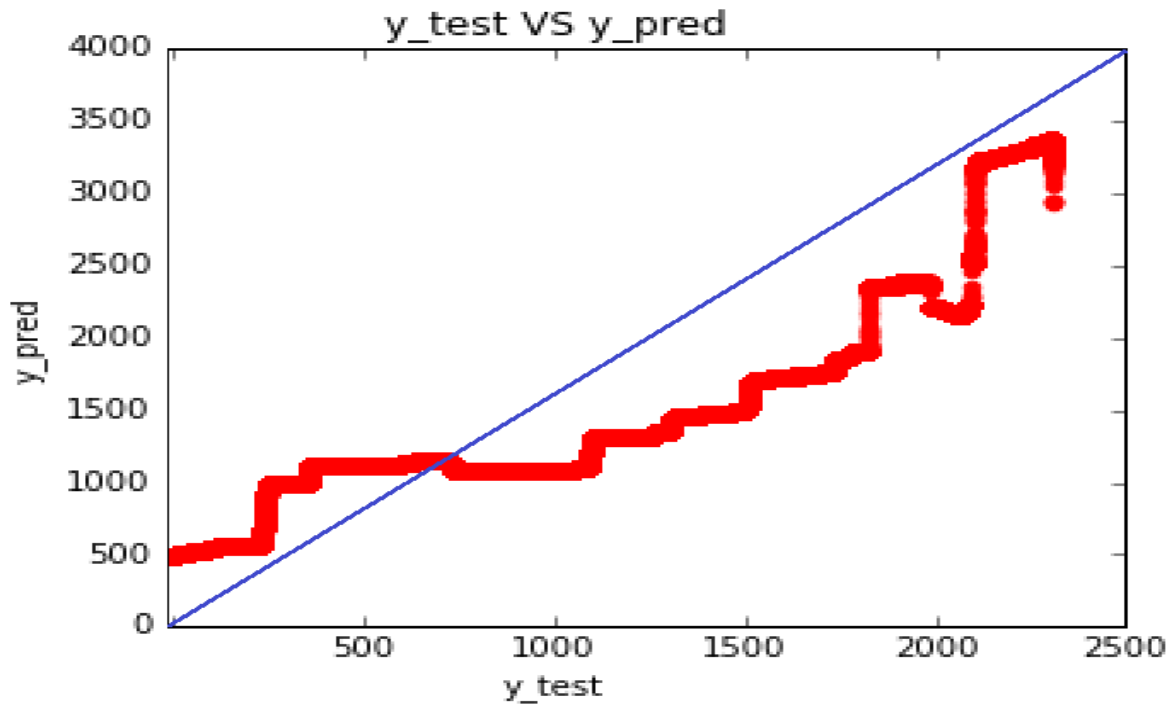

**Figure 9:** Prediction values vs actual values of remaining flight time using Random Forest model.

### 5.5.2. Neural Network

Neural network models are powerful predictive models which can capture complex relationships from a large dataset well. Therefore, a neural network is considered for building a predictive model. As a neural network model can perform variable selection on its own by assigning appropriate weights, the first five principal components which explain 99.999% of the variance are used as input predictors for this model. The number of hidden layers and nodes are varied to identify a better fitting model.

A total of four neural network model configurations were considered. The first model with a single hidden layer of 3 nodes was built. Single hidden layers with three nodes are chosen as a starting network as a simple network configuration. The neural network model with weights and bias on this neural network model after training is shown in Figure 10-(a) and the goodness of fit of prediction on a new battery-flight test data with this model is shown in Figure 10-(b). The predicted values are a little off from the actual values. It can be observed that this model under-predicts most of the time with a small over-prediction towards the end. The MSE on prediction is 35763 for this model. The model appears to be not complex enough to capture the relationship between the predictor variables.

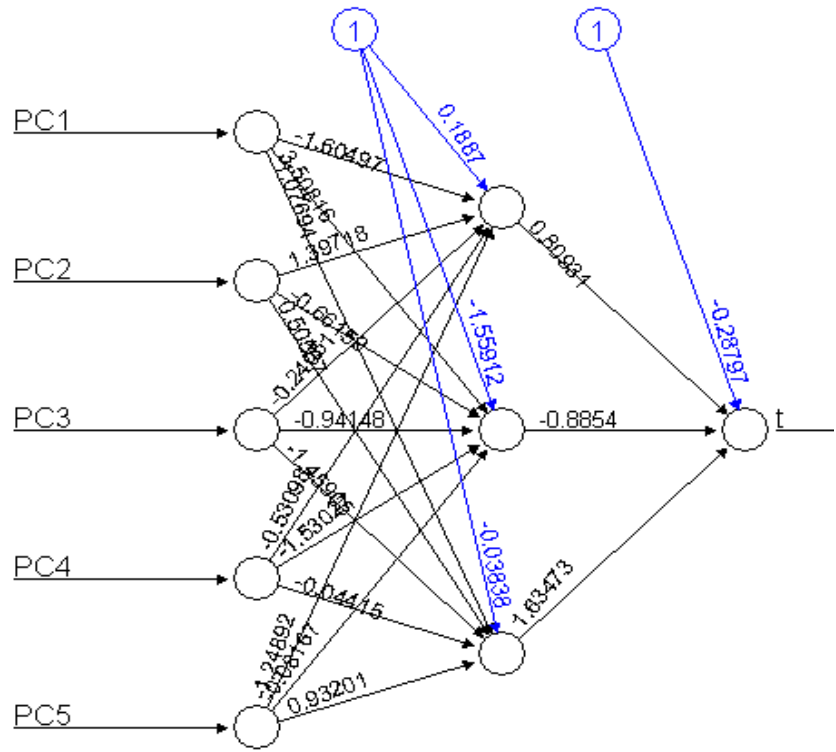

**Figure 10-(a):** Weights and Bias in Neural Net of single hidden layer of 3 nodes

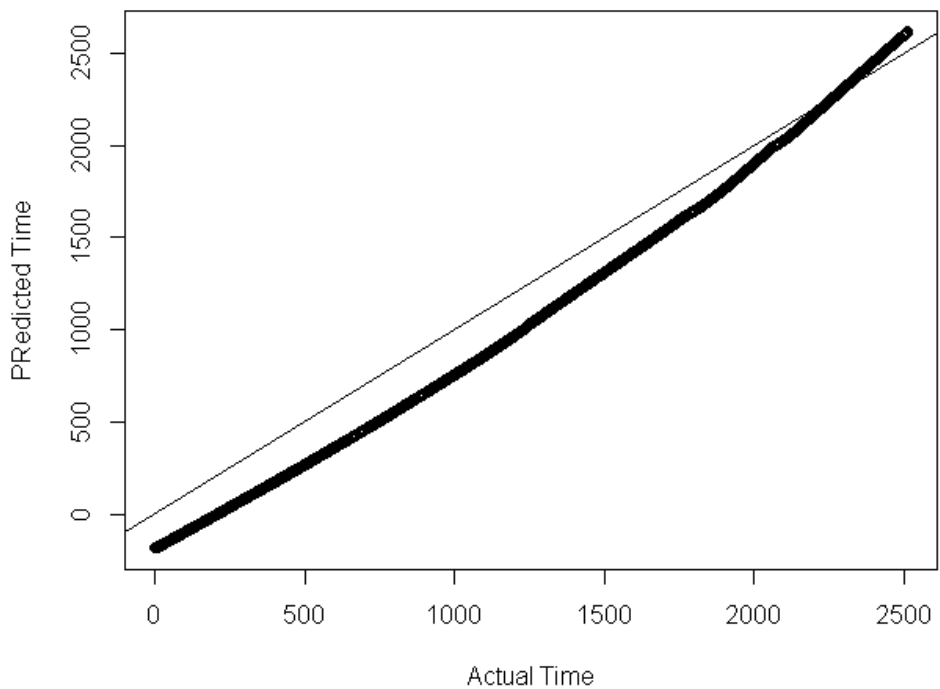

**Figure 10-(b):** Goodness of fit for Neural Net of single hidden layer of 3 nodes

Furthermore, a second model with a single hidden layer of 5 nodes was considered. The hope was the additional nodes in the model would help capture complex relationships and interactions between predictors well. The higher nodes would add additional weight and bias parameters to the models and therefore more complex models with more computational time for training. Figure 11-(a) shows the weights and bias from the trained neural net and Figure 11-(b) shows the goodness of fit prediction plot for this neural network model. It can be observed that prediction accuracy is increased with 5 nodes model significantly compared to the previous model of 3 nodes. The predicted time is following the actual time very closely all the time. Figure 11-(b) clearly illustrates the significant reduction in prediction error and well-matching of predicted values of time to actual values. The MSE of the prediction is significantly reduced to 2717. Clearly, the addition of 2 nodes to the neural network helped reduced MSE by 92%. The higher complexity of the model with the addition of 2 nodes is well justified by such a significant reduction in the MSE.

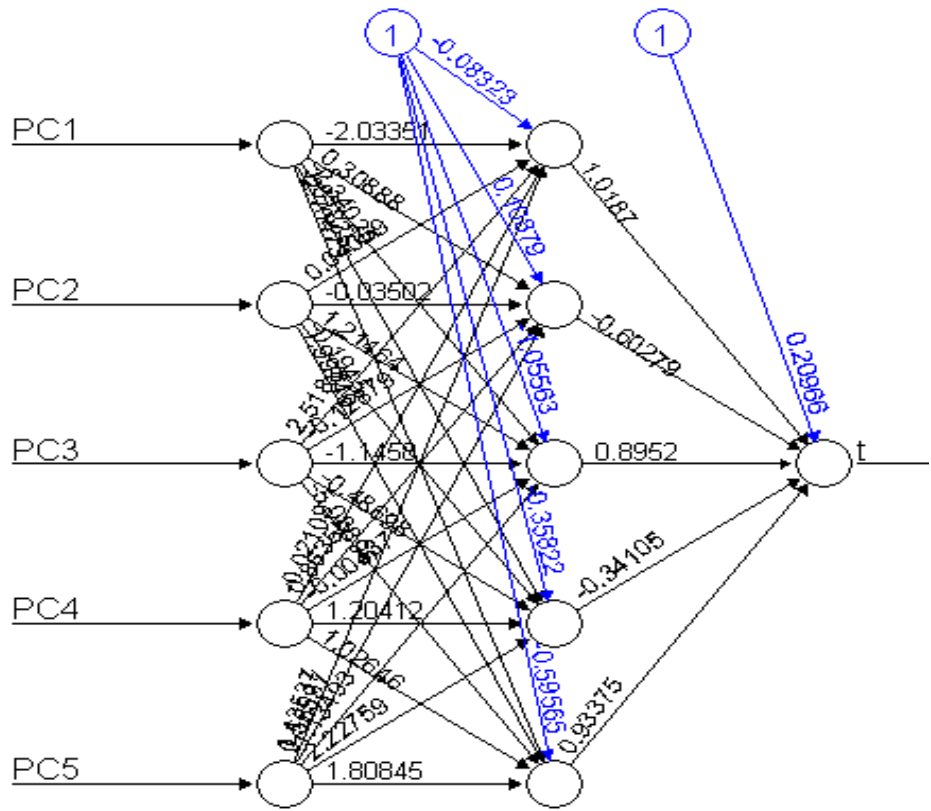

**Figure 11-(a):** Weights and Bias in Neural Net of single hidden layer of 5 nodes

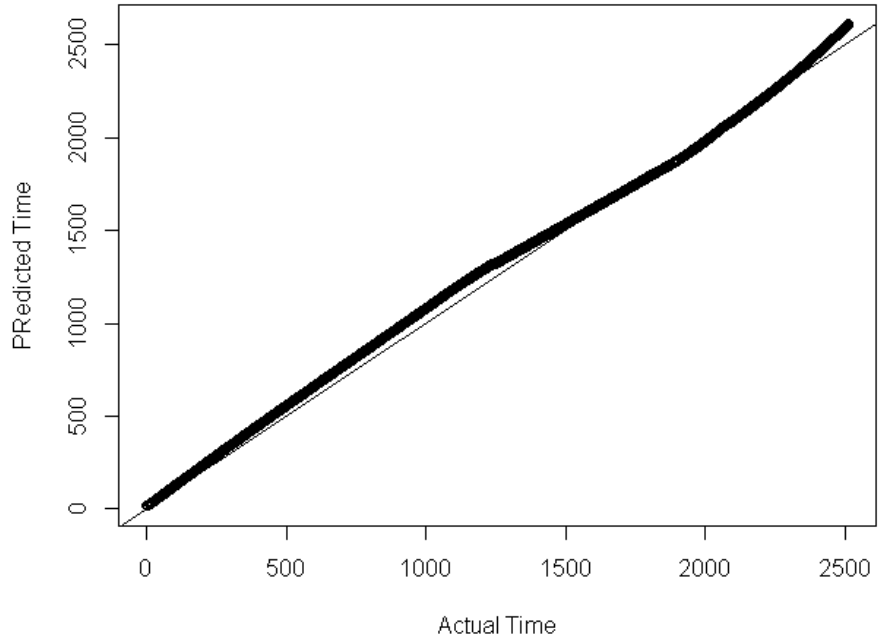

**Figure 11-(b):** Goodness of fit for Neural Net of a single hidden layer of 5 nodes

Further, complex deep learning neural net configurations were also explored. The neural net configurations with two hidden layers with the first layer of 5 nodes and the second layer of 1 node are trained. The model configuration with values of weights and biases after training is shown in Figure12-(a). Also, the goodness of fit with two hidden layer models with 5 nodes in the first layer and a node in the single layer is shown in Figure 12-(b). It is observed that prediction accuracy decreases for this model compared to the single-layer network of 5 nodes. The MSE value on test data was 11896. Clearly, 338% higher MSE was observed over the MSE of a single layer of 5 nodes.

Although two hidden layer model showed poor fit with higher MSE on test data, another two hidden model with more nodes with 5 nodes in the first layer and 3 nodes in the second layer is explored to confirm and understand the possibility of improvement of model fitting with additional nodes in the second layer. Figure 13-(a) shows the network configuration with weights and bias values after training and Figure13-(b) shows the goodness of fit with this network configuration on test data. The model with the first layer of 5 nodes and the second layer of 3 nodes also showed poor prediction performance with an MSE of 15709. It is observed that prediction is significantly worse with 2-layer models than with the single-layer 5-node model. Although training error decreased with the complex neural network, prediction error on test data was observed to be the least with the model having a single layer of 5 nodes. Therefore, a single-layer model with 5 nodes is the best model with the neural network modeling approach.

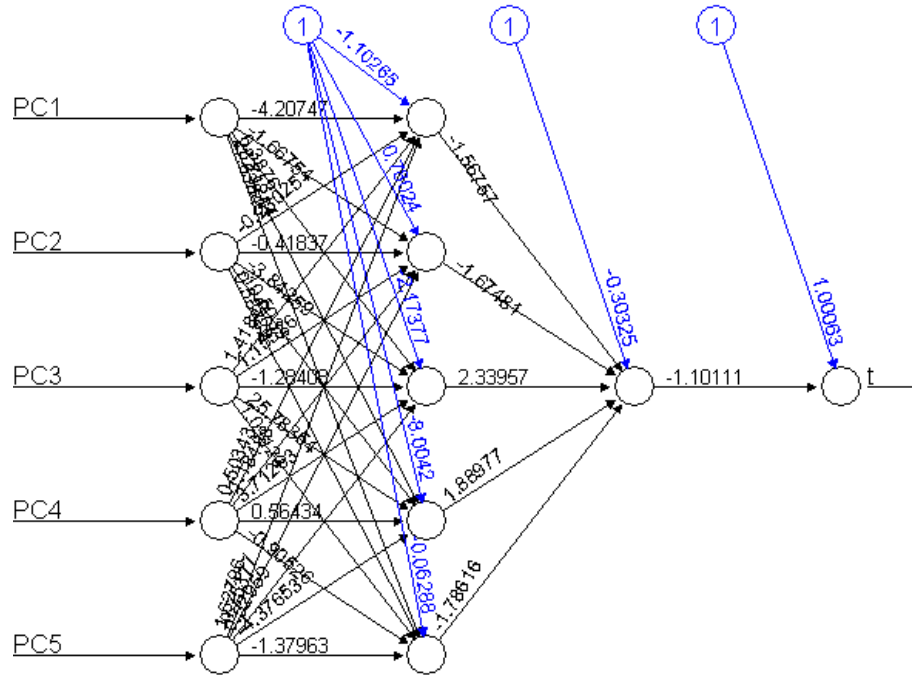

**Figure 12-(a):** Weights and Bias in Neural Net of 2 Layers with 5 nodes in the first layer and 1 node in the second layer.

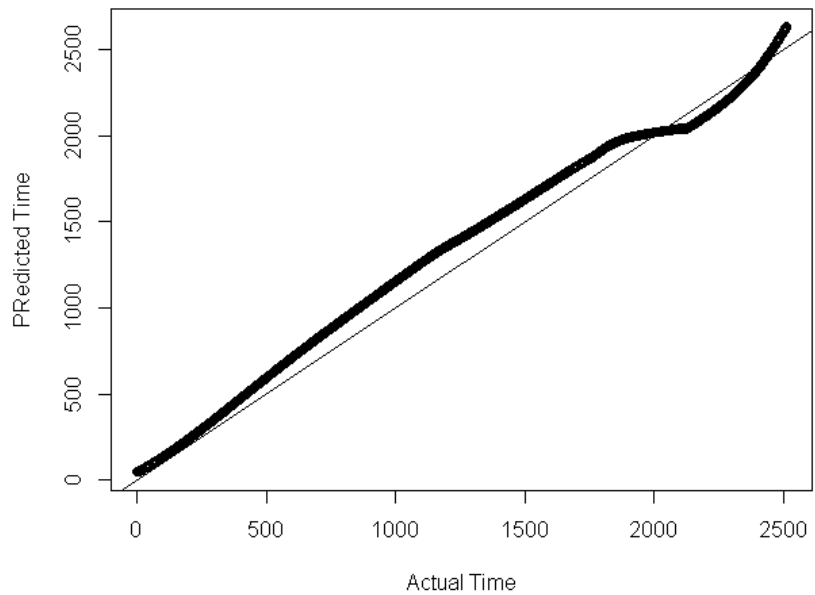

**Figure 12-(b):** Goodness of fit for Neural Net of 2 Layers with 5 nodes in the first layer and 3 nodes in the second layer.

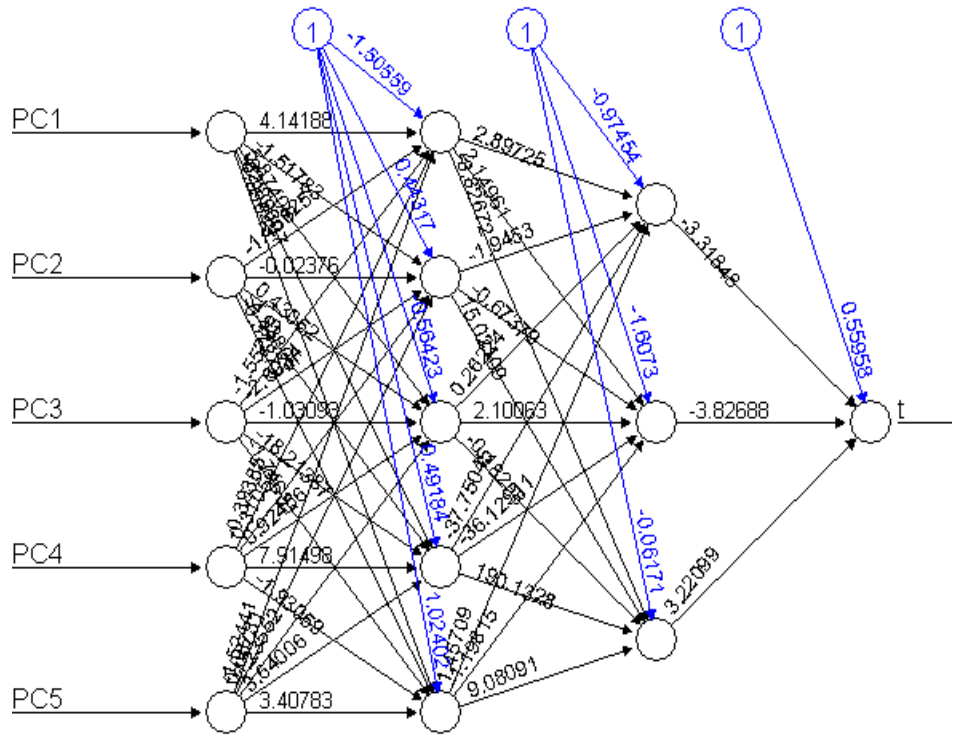

**Figure 13-(a):** Weights and Bias in Neural Net of 2 Layers with 5 nodes in the first layer and 3 nodes in the second layer.

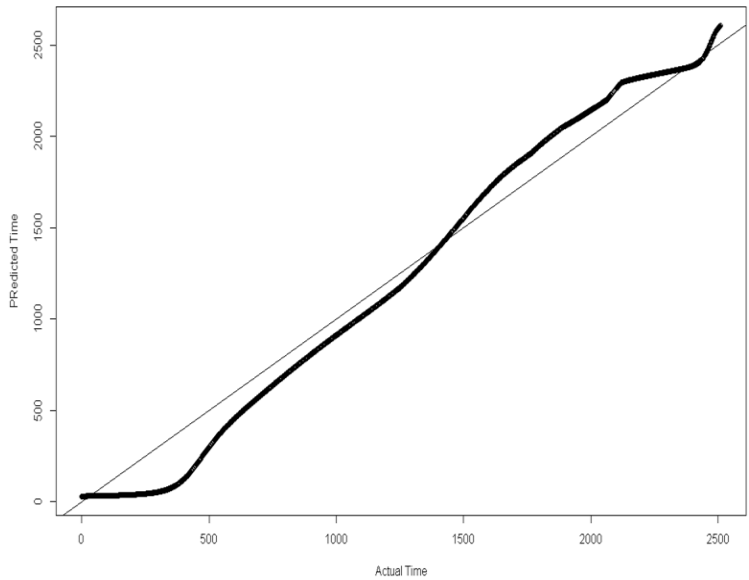

**Figure 13-(b):** Goodness of fit for Neural Net of 2 Layers with 5 nodes in the first layer and 3 nodes in the second layer.

### 5.5.3. Model Comparison

The prediction performance of Random Forest and Neural Network is compared in terms of prediction accuracy, computational time, and robustness of the model in handling higher dimensional data. Prediction accuracy is measured in terms of MSE on test data for model comparison. The approximate time required for training the model was considered as a measure of computational time. The robustness of the model was considered based on the model's ability to handle high-dimensional data without the need for dimensionality reduction. Table 1. provides performance measures of the Random Forest model and various neural network models considered in this study. Prediction MSE for the best Neural Network model was 2717 whereas Prediction MSE for the Random Forest model was 27431. Clearly, the neural network modeling approach showed a significantly smaller prediction error. However, the computational time required for training neural networks was significantly higher (about 3 times) even with a reduced data set. Additionally, neural networks can handle feature selection whereas Random Forest requires a feature selection process before model building. Therefore, the top two principal components were used for random forest whereas the top five principal components were considered for neural network models. Neural network models are more robust in handling high-dimensional data than random forest models. Just based on MSE of prediction on test data, a neural network model with a single hidden layer of 5 nodes is observed to be the best model for prediction of battery lifetime.

**Table 1:** Model comparison for Random Forest and various neural networks (NN) models on battery dataset.

| Model | Test MSE | Computational Time | Robustness |
|---|---|---|---|
| Random Forest | 27431 | ~ 10 minutes | Low |
| NN-Single Layer 3 nodes | 35763 | ~ 30 minutes | High |
| NN- Single Layer 5 nodes | 2717 | ~ 35 minutes | High |
| NN- Double Layer (5,1) nodes | 11896 | ~ 50 minutes | High |
| NN- Double Layer (5,3) nodes | 15709 | ~ 55 minutes | High |

# 6. Conclusions

Data-driven modeling approach was explored to predict the Lithium-ion battery lifetime based on battery data for electric-powered aircraft, using the measures of aircraft state and battery SOCs over its usage. It was observed that feature engineering was necessary for predictive model building. Features that provide battery utilization information over its operation were needed. Therefore, new features were generated from the original instantaneous battery utilization information to provide battery utilization over its past operation using the area under the curve measures. Additionally, many features had a high correlation to other features therefore dimensionality reduction was explored using principal component analysis.

After feature engineering and principal component analysis, two predictive models were developed based on Random Forest and Neural Network models. For the random forest model ensemble model based on 50 decision trees was built using training data. The prediction MSE on test data was considered as a performance metric for model comparison. Moreover, the goodness of fit plot was considered to see visually how model predictions fit to the actual lifetime data. For the exploration of neural network models, four neural network configurations were considered in this study. These four model configurations include a single layer with 3 nodes, a single layer with 5 nodes, double layers with 5 nodes in the first layer and 1 node in the second layer, and a double layer with 5 nodes in the first layer and 3 nodes in the second layer. All four neural network models were trained to evaluate weights and bias parameters of the model using 5 principal components as inputs and full training data. It was observed that the neural network model with a signal layer with 5 nodes is the best model for MSE performance metric on test data. Clearly, too less nodes underfit or too excess nodes in neural network results in overfitting. The MSE on training data was consistently decreasing with higher nodes, however, after optimal node configuration, MSE on test data was increasing. This clearly highlights optimal node configuration should be decided based on MSE on test data.

Overall, the best neural network model had an MSE of 2717 whereas the prediction MSE for the Random Forest model was 27431 on the test data. Clearly based on MSE, the neural network model performed better and yielded better predictions of battery lifetime. Moreover, the computational time for training neural network models was 3 times higher than random forest models. This higher computational time is justified as over 90% reduction in MSE is observed with the best neural network model. Furthermore, the neural network model was more robust as it could handle higher dimensional data without the need for extensive dimensional reduction. The model comparison clearly highlights the accuracy and robustness of neural network models compared to Random Forest models for battery lifetime prediction. Therefore, it is possible to predict battery lifetime using real-world battery data using a data-driven modelling approach with reasonable accuracy.

## 7. Future work

This study primarily considered the two most versatile machine learning models, many more models can be explored beyond Random Forest and Neural network models. Moreover, battery lifetime prediction for different applications can be explored using a data-driven modelling approach. One important aspect for the data-driven modeling approach for battery lifetime prediction was feature engineering and high correlation of predictor variables. It is important to measure appropriate data on battery for accurate lifetime prediction. Therefore, additional analysis can be explored to identify methods to reduce data dimensionality at the time of battery data measurement itself so that more informative variables were included as part of battery management system measurements.

Further research and development of other models such as functional regression, hybrid models using physics-based model and machine learning models for more accurate and robust battery lifetime estimate. Additionally, quantification of uncertainty measures for the prediction could be explored to identify data-driven models with lower uncertainty of predictions. This is highly desired for critical applications where battery lifetime prediction needs to be with lower uncertainty. Overall, this study explores a small aspect of possible modeling approaches for battery lifetime prediction. Many more approaches still remain to be explored for more accurate, computationally effective and reliable battery lifetime prediction.